\documentclass[11pt]{JHEP3}

\usepackage{amsmath}
\usepackage{amssymb}
\usepackage{slashed}

\DeclareMathOperator{\tr}{tr}   
\newcommand{\imu}{{\rm i}}
\newcommand{\LL}{{\lambda}}
\newcommand{\OO}{{\Omega}}
\renewcommand{\SS}{{\Sigma}}
\newcommand{\oo}{{\omega}}
\newcommand{\TT}{{\theta}}
\newcommand{\Ei}{{\rm Ei}}

\title{Computation of the Chiral Anomaly in the Bulk Quantization}

\author{Thomas Klose and Sorin Marculescu \\ Fachbereich Physik, Universit\"{a}t Siegen, D-57068 Siegen, Germany \\ E-mail: \email{klose@cip-lx1.physik.uni-siegen.de}, \email{marculescu@aleph.physik.uni-siegen.de}}

\abstract{The bulk quantization method is used for regularizing a conventional four dimensional theory of massless fermions coupled to an external non-Abelian gauge field and for subsequently evaluating the associated Ward identity. As a result one obtains the well-known chiral anomaly.}


\preprint{hep-th/0202102}

\begin{document}


\section{Introduction}

\vspace{0.5cm}

\noindent Recently, a new version of stochastic quantization \cite{damhue}, \cite{nam} called bulk quantization has 
been proposed \cite{bauzwa1}, \cite{grabauzwa}, \cite{bauzwa2}. The approach seems to be especially convenient for 
gauge theories (including matter).

Bulk quantization is expressed in terms of a $\rm{4} + \rm{1}$-dimensional stochastic action dynamically equivalent to the Langevin equation. Of course, for spinor matter the Langevin equation has a non-trivial kernel \cite{sak}. The full stochastic action for QCD type theories and the associated propagators are written in superfield form in reference \cite{dipl}. Translation invariance is extended to the fifth coordinate, assuming therefore trivial initial conditions. By adding to the action above a certain topological term a new symmetry called (fictitious) time reversal appears. On the basis of the translational and time reversal invariances it is possible to show that the correlations of physical observables in the bulk quantization and in the conventional Euclidean four-dimensional theory are the same. It follows further that if the four-dimensional theory has a global invariance the associated Ward identities hold on a fictitious time-slice of the five-dimensional theory.

This property does not apply if the four-dimensional Ward identity is anomalous \cite{bauzwa2}. This means that some  correlation functions of the current divergence do not vanish. We shall show that the equal time limit of such a correlation coincides -- as expected -- with the result of the conventional Euclidean theory. The fact that the equal time limit  of the triangle correlator is ambiguous \cite{grabauzwa} plays no role, because additional requirements like symmetry considerations, consistency conditions or vector-current conservation provide a unique anomaly.
 
While the full merits of stochastic quantization are revealed in a non-per\-tur\-ba\-tive analyse, it is customary to  check that perturbation calculations lead to standard results. Such a test has been performed long time ago for pure Yang-Mills theories at one and two-loop level \cite{oka}, \cite{marschokazhe}. Since we are not aware of similar  computations in the spinor sector of gauge theories we consider the present work also as a tentative to fill this gap, too. Nevertheless there were attempts to regularize stochastic quantization of gauge field theories with fictitious time as cut-off parameter \cite{breit}. For instance, the consistent anomaly was computed in this regularization scheme by using a variant of the Fujikawa method \cite{tzani}. More recently, a mathematically rigorous proof of the Atiyah-Singer index theorem has been achieved on the basis of a graded stochastic calculus \cite{rogers}.
 
The paper is organized as follows: In section \ref{sec:effective_action} the regularization of the effective action  of the conventional theory of massless fermions coupled to an external non-Abelian gauge field is performed in the  framework of bulk quantization. It appears that the divergent part of the effective action is gauge invariant  implying a finite BRS variation. By an appropriate extension of the BRS operation we are able in section \ref{sec:ward_identity} to regularize the Ward identity and to compute the chiral anomaly. Some technical details are collected in the appendix.


\section{Effective Action and Its Divergent Part} \label{sec:effective_action}

\vspace{0.5cm}

Consider the stochastic action of a system of free fermions: 
\begin{eqnarray}
{\cal S}_0 & \equiv & \int d_4x d\tau \biggl[ 2 \LL b^\dag ( x, \tau ) \slashed{\partial} b ( x, \tau ) + b^\dag ( x, 
\tau ) \biggl( \frac{\partial}{\partial \tau} - \LL \square \biggr) q ( x, \tau ) \nonumber \\ 
           &        & \mbox{} + q^\dag ( x, \tau ) \biggl( \stackrel{\leftarrow}{\frac{\partial}{\partial \tau}} - 
\LL \stackrel{\leftarrow}{\square} \biggr) b ( x, \tau ) \biggr] \; ,
\end{eqnarray}
where $q ( x, \tau )$, $q^\dag ( x, \tau )$ denote the quark fields and $b ( x, \tau )$, $b^\dag ( x, \tau )$ the auxiliary fields which represent the Lagrange multipliers for the spinor Langevin equations. For simplicity we integrated out the (commuting) spinor ghosts and antighosts. Throughout this work the spinor fields are in the representation of a compact Lie gauge group with anti-Hermitean generators $T^a$. The parameter $\LL$ has no direct physical meaning and could be set equal to one. However, we keep it explicitly because it turns out to be a convenient book keeping quantity for the expansion about the equal time limit. Further notations and conventions are explained in the appendix.

The correlations of the left chiral current
\begin{equation}
j_{\mu}^a ( x, \tau ) \equiv - q^\dag ( x, \tau ) T^{a} {\gamma}_{\mu} P_L q ( x, \tau ) \label{eqn:left_chiral_current}
\end{equation}
are obtained from the effective stochastic action
\begin{eqnarray} 
W ( A ) & \equiv & \ln \int [ dq dq^\dag db db^\dag ] ( x, \tau ) \exp \left[ - {\cal S}_0 + \int d_4x d \tau j_{\mu}^a ( x, \tau ) A_{\mu}^a ( x, \tau ) \right] \nonumber \\
        &    =   & \ln \int [ dq dq^\dag db db^\dag ] ( x, \tau ) \nonumber \\
        &        & \mbox{} \times \exp \left[ - {\cal S}_0 - \int d_4x d \tau q^\dag ( x, \tau ) \slashed{A} ( x, 
\tau ) P_L q ( x, \tau ) \right] \; . \label{eqn:stochastic_effective_action}
\end{eqnarray}    
In the last line a Lie algebra valued form has been used for the vector field $A_{\mu} ( x, \tau ) \equiv A_{\mu}^{a} ( x, \tau ) T^{a}$. The classical field $A_{\mu}^{a} ( x, \tau )$ acts as a source for the left chiral composite object \eqref{eqn:left_chiral_current}. It has mass dimension $3$ and is related to the external non-Abelian gauge field $a_{\mu}^{a} ( x )$ by
\begin{equation}
A_{\mu}^{a} ( x, \tau ) =  a_{\mu}^{a} ( x ) \delta ( \tau ) \; , \label{eqn:identification_gauge_field} 
\end{equation}
but for many purposes it is sufficient to assume that the source $A_{\mu}^{a} ( x, \tau )$ is a homogeneous function 
of degree $-1$:
\begin{equation}
A_{\mu}^{a} ( x, \rho \tau ) =  {\rho}^{- 1} A_{\mu}^{a} ( x, \tau ) \; . \label{eqn:homogeneity_gauge_field}
\end{equation}
       
The effective action of the conventional four-dimensional theory of massless fermions coupled to an external non-Abelian gauge field $a_{\mu}^{a} ( x )$ is given by 
\begin{equation} 
W (a) = \ln \int [ d \psi d{\psi}^\dag ] ( x ) \exp \left\{ - \int d_4 x {\psi}^\dag ( x ) \left[ \slashed{\partial} + \slashed{a} ( x ) P_L \right] \psi ( x ) \right\} \; . \label{eqn:conventional_effective_action} 
\end{equation}
Here we used the possibility \cite{wei} of writing the gauge coupling of a general massless fermionic matter in left chiral form.

The correlations obtained from \eqref{eqn:conventional_effective_action} have ultraviolet divergences. Since the source term in \eqref{eqn:stochastic_effective_action} becomes the interaction of the conventional model \eqref{eqn:conventional_effective_action} when the identification \eqref{eqn:identification_gauge_field} is made, one can use the effective action in the bulk quantization to regularize the effective action of the conventional theory. 

We proceed by expanding \eqref{eqn:stochastic_effective_action} according to the number of external lines
\begin{equation} 
W ( A ) = \sum_{n=2}^{\infty} W_n ( A ) \; , \label{eqn:expansion_effective_action}
\end{equation}
where the (divergent) contribution of the tadpole diagram
\begin{equation}
W_1 ( A ) \equiv - \imu \int d_4 x d \tau \int \frac{d_4 p}{( 2 \pi)^4} \tr \left[ \frac{\slashed{p}}{p^2} \slashed{A} ( x, \tau ) P_L \right] \; \label{eqn:tadpole} 
\end{equation}  
has been neglected in \eqref{eqn:expansion_effective_action} because the integrand of \eqref{eqn:tadpole} is odd in momentum. (This argument works also when the gauge group is not semi-simple, i.~e. $\tr T^a \neq 0$.) 

The contribution of a loop with $n \geq 2$ external lines is given by
\begin{eqnarray}
W_n ( A ) & \equiv & \int \prod_{i=1}^{n} d{\tau}_i \int \prod_{i=1}^n \frac{d_4 k_i}{( 2 \pi )^4} {\cal W}_{{\mu}_1 \ldots {\mu}_n}^{(n)} \left( {\tau}_1, \ldots , {\tau}_n ; k_1, \ldots , k_{n-1} \right) \nonumber \\
          &        & \mbox{} \times ( 2 \pi )^4 {\delta}_4 \left( \sum_{i=1}^n k_i \right)\tr \left[ A_{{\mu}_1} ( - k_1, {\tau}_1 ) \ldots A_{{\mu}_n} ( - k_n, {\tau}_n ) \right] \; , \quad \label{eqn:expansion_terms}
\end{eqnarray}
where the loop integral has the form
\begin{eqnarray}
{\cal W}_{{\mu}_1 \ldots {\mu}_n}^{(n)} & \equiv & {\cal W}_{{\mu}_1 \ldots {\mu}_n}^{(n)} \left( {\tau}_1, \ldots , {\tau}_n ; k_1, \ldots , k_{n-1} \right) \nonumber \\
                                        &   =    & - \frac{{\imu}^n}{n} \int \frac{d_4 p}{( 2 \pi )^4} \exp \left( - \LL \sum_{i=1}^n p_i^2 t_i \right) \tr \left( \frac{\slashed{p}_1}{p_1^2}{\gamma}_{{\mu}_1} \ldots \frac{\slashed{p}_n}{p_n^2}{\gamma}_{{\mu}_n} P_L \right) \qquad\;\; \label{eqn:expansion_terms_integrand}
\end{eqnarray}
and where the Fourier transform is defined by
\begin{equation}
A_{\mu} ( k, \tau ) \equiv \int d_4 x e^{- \imu k x} A_{\mu} ( x, \tau ) \; .
\end{equation}
The following abbreviations were used in \eqref{eqn:expansion_terms_integrand}
\begin{eqnarray}
& p_1 \equiv p \; ;  & p_i \equiv p + \sum_{j=1}^{i-1} k_j \; \quad \mbox{for $i = 2, \ldots , n$} \; ; \label{eqn:abbrev_1} \\ 
& t_1 \equiv  | {\tau}_n - {\tau}_1 | \; ; & t_i \equiv | {\tau}_i - {\tau}_{i-1} | \; \quad \mbox{for $i = 2, \ldots , n$} \; .
\end{eqnarray}
Occasionally, we write
\begin{equation}
k \equiv \sum_{j = 1}^{n - 1} k_j \; ; \qquad \; l \equiv k + k_{n - 1} \; ; \qquad \; t \equiv \sum_{i = 1}^n t_i \; . \label{eqn:abbrev_2}
\end{equation}

Note that in \eqref{eqn:expansion_terms_integrand} the parameter $\LL$ multiplies all time differences $t_i$. The limit 
\begin{equation}
\LL \to 0  \nonumber
\end{equation}
therefore has the same effect as making the identification \eqref{eqn:identification_gauge_field} and in addition it is a convenient method to separate and classify the divergences.
 
For $n \geq 5$ the integral \eqref{eqn:expansion_terms_integrand} converges at $\LL = 0$, such that the identification \eqref{eqn:identification_gauge_field} can be made in \eqref{eqn:expansion_terms}. The infinite part of the effective action consists of divergent contributions of loops with two, three and four external lines
\begin{equation}
W^{div} ( a ) = \sum_{n=2}^{4} W_n^{div} ( a ) \; . 
\end{equation}

The two-leg contribution $W_2 ( A )$ splits into two parts, a transversal part  $W_{2 \; \perp} ( A )$ and a remaining $W_{2 \; \parallel} ( A )$ which turns out to be zero. In order to see this let us write the remaining part in the form
\begin{eqnarray}
W_{2 \; \parallel} ( A ) & = & \int d{\tau}_1 d{\tau}_2 \int \frac{d_4 k}{( 2 \pi )^4} \; {\cal W}^{( 2 \; \parallel )} \left( {\tau}_1, {\tau}_2 ; k \right) \nonumber \\
                         &   & \mbox{} \times \tr \left[ A_{\mu} ( - k, {\tau}_1 )  A_{\mu} ( k, {\tau}_2 ) \right] \; , \label{eqn:w2parallel}
\end{eqnarray}
where we introduced
\begin{equation}
{\cal W}^{( 2 \; \parallel )} =  - \int \frac{d_4 p}{( 2 \pi )^4} \exp \left\{ - \LL \frac{t}{2} \left[ p^2 + ( p + k 
)^2 \right] \right\} \tr \left[ \frac{\slashed{p}}{p^2}\frac{\slashed{p} + \slashed{k}}{( p + k )^2}  P_L \right]   
\end{equation} 
with $t\equiv 2 |{\tau}_2 - {\tau}_1 |$. We use now a mixed parameterization, a Feynman parameter $u$ for combining denominators and a parameter $\beta$ for exponentiating the resulting denominator. By an appropriate shift in the  loop momentum $p$ and subsequent symmetric integration one obtains 
\begin{eqnarray}
{\cal W}^{( 2 \; \parallel )} & = & - \frac{1}{( 4 \pi )^2} \int_0^1 du \int_0^{\infty} d \beta \frac{\beta}{B^3} \exp \left\{ - \frac{k^2}{B} \left( \frac{\LL t}{2} + \beta u \right) \left[ \frac{\LL t}{2} + \beta ( 1 - u ) \right] \right\} \nonumber \\
                              &   & \mbox{} \times \left\{ 1 + \frac{k^2}{B} \left( \frac{\LL t}{2} + \beta u \right) \left[ \frac{\LL t}{2} + \beta ( 1 - u ) \right] \right\} \; , \label{eqn:w2parallel_integrand}
\end{eqnarray}
where $B \equiv \LL t + \beta$. (The required formulae are given in the appendix.) Following the trick of reference \cite{itzzub} one writes \eqref{eqn:w2parallel_integrand} as
\begin{eqnarray}
{\cal W}^{( 2 \; \parallel )} & = & \frac{\partial}{\partial \rho} \frac{1}{( 4 \pi)^2}  \int_0^1 d u \int_0^{\infty} d \beta \frac{\beta}{B^3 \rho} \nonumber \\ 
                              &   & \mbox{} \times \left. \exp \left\{ - \rho \frac{k^2}{B} \left( \frac{\LL t}{2} + \beta u \right) \left[ \frac{\LL t}{2} + \beta ( 1 - u ) \right] \right\} \right|_ {\rho=1} \; . \label{eqn:w2parallel_integrand_rho}
\end{eqnarray}
Inserting \eqref{eqn:w2parallel_integrand_rho} into \eqref{eqn:w2parallel}, scaling both fictitious time variables by the same $\rho$ and using the homogeneity property \eqref{eqn:homogeneity_gauge_field} of the current sources one  can remove any $\rho$ dependence from the integral. Hence \eqref{eqn:w2parallel} vanishes.

The transversal part is given by
\begin{eqnarray}
W_{2 \; \perp} & = & \int d{\tau}_1 d{\tau}_2 \int \frac{d_4 k}{( 2 \pi )^4} \left( k_{\mu}k_{\nu} - {\delta}_{\mu \nu} k^2 \right) {\cal W}^{( 2 \; \perp )}\left( {\tau}_1, {\tau}_2 ; k \right) \nonumber \\   
               &   & \mbox{} \times \tr \left[ A_{\mu} ( - k, {\tau}_1 ) A_{\nu} ( k, {\tau}_2 ) \right] \; , \label{eqn:w2transversal} 
\end{eqnarray}
where
\begin{eqnarray}
{\cal W}^{( 2 \; \perp )} & = & - \frac{1}{8 {\pi}^2} \int_0^1 du \int_0^{\infty} d \beta \frac{\beta}{B^4} \left( \frac{\LL t}{2} + \beta u \right) \left[ \frac{\LL t}{2} + \beta ( 1 - u ) \right] \nonumber \\
                          &   & \mbox{} \times \exp \left\{ - \frac{k^2}{B} \left( \frac{\LL t}{2} + \beta u \right) \left[ \frac{\LL t}{2} + \beta ( 1 - u ) \right] \right\}
\end{eqnarray}
diverges logarithmically in the limit $t = 0$ or $\LL = 0$. Since we are interested just in such behavior we can discard terms of order $\LL t$ and obtain
\begin{eqnarray}
{\cal W}^{( 2 \; \perp )} & = & - \frac{1}{8 {\pi}^2} \int_0^1 du \int_0^{\infty} d \beta \frac{{\beta}^3}{B^4} u(1-u) \exp [ - B k^2 u ( 1 - u ) ] \nonumber \\
                          &   & \mbox{} \times  \left[ 1 + {\cal O} ( \LL t ) \right] \; . \label{eqn:w2transversal_integrand}
\end{eqnarray}
We are now in the position to apply the results stated in the appendix. Since $P ( u, 0 ) = u ( 1 - u )$ we use \eqref{eqn:theorem_integral_nonzero} and find
\begin{equation}
{\cal W}^{( 2 \; \perp )} = - \frac{1}{48 {\pi}^2} \ln \left( \LL t k^2 \right) + {\cal O} ( \LL t ) \; .
\end{equation} 

In order to separate the ultraviolet divergence from the infrared one, we introduce a mass scale $\mu$. Recall that the classical theory is massless, so there is no natural mass scale available. We use now the expansion
\begin{equation} 
\ln ( \LL t k^2 ) = \ln ( \LL t {\mu}^2 ) + \frac{k^2 - {\mu}^2}{{\mu}^2} + \ldots \label{eqn:expansion_logarithm} 
\end{equation}
in \eqref{eqn:w2transversal_integrand} and subsequently in \eqref{eqn:w2transversal}. By making the identification \eqref{eqn:identification_gauge_field} one can see that the divergence of $W_2 ( A )$ comes entirely from the logarithm and is given by 
\begin{eqnarray} 
W^{div}_2 ( A ) & = & - \frac{1}{48 {\pi}^2} \int d{\tau}_1 d{\tau}_2 \ln ( \LL t {\mu}^2 ) \nonumber \\
                &   & \mbox{} \times \int d_4 x \tr \left[ A_{\mu} ( x, {\tau}_1 ) \left( {\partial}_{\mu} {\partial}_{\nu} - {\delta}_{\mu \nu} \square \right) A_{\nu} ( x, {\tau}_2 ) \right] \; .
\end{eqnarray}
The $\tau$ integration can be restricted to the interior of a small square around the origin
\begin{equation} 
2 | {\tau}_1 \pm {\tau}_2 | \leq T \ll \frac{1}{{\mu}^2} \; . \label{eqn:tau_integration_domain_2}
\end{equation}
Using a mean value theorem one can extract the logarithmic term in front of the $\tau$ integral and perform the $\tau$ integration according to \eqref{eqn:identification_gauge_field}. The result is:
\begin{equation} 
W^{div}_2 ( a ) = - \frac{\ln ( \LL T {\mu}^2 )}{48 {\pi}^2} \int d_4 x \tr \left[ a_{\mu} ( x ) \left( {\partial}_{\mu} {\partial}_{\nu} - {\delta}_{\mu \nu} \square \right) a_{\nu} ( x ) \right] \; . \label{eqn:w2_result}
\end{equation}

The loop diagram with three external lines is given by
\begin{eqnarray} 
W_3 ( A ) & = & \int d{\tau}_1 d{\tau}_2 d{\tau}_3 \int \frac{d_4 k_1 d_4 k_2}{( 2 \pi )^8} {\cal W}^{( 3 )}_{\mu \nu \rho} \left( {\tau}_1,{\tau}_2, {\tau}_3 ; k_1, k_2 \right) \nonumber \\
          &   & \mbox{} \times \tr \left[ A_{\mu} ( - k_1, {\tau}_1 ) A_{\nu} ( - k_2, {\tau}_2 ) A_{\rho} ( k, {\tau}_3 ) \right] \; , \label{eqn:w3}
\end{eqnarray}
where
\begin{eqnarray} 
{\cal W}^{( 3 )}_{\mu \nu \rho} & = & \frac{\imu}{3} \tr \left( {\gamma}_{\alpha} {\gamma}_{\mu}{\gamma}_{\beta} {\gamma}_{\nu} {\gamma}_{\gamma} {\gamma}_{\rho} P_L \right) \int \frac{d_4 p}{( 2 \pi )^4} \frac{p_{\alpha} ( p + k_1 )_{\beta} ( p+ k )_{\gamma}}{p^2 ( p + k_1 )^2 ( p + k )^2} \nonumber \\
                                &   & \mbox{} \times \exp \left\{ - \LL \left[ p^2 t_1 + ( p + k_1 )^2 t_2 +  ( p + k )^2 t_3 \right] \right\} \; .  
\end{eqnarray}
Since we now have two independent external momenta $k_1$ and $k_2$ it is convenient to make an expansion of the integrand in order to have two independent denominators, say $p^2$ and $( p + k_1 )^2$:
\begin{equation}
\frac{p_{\alpha} ( p + k_1 )_{\beta} ( p+ k )_{\gamma}}{p^2 ( p + k_1 )^2 ( p + k )^2} =  \frac{p_{\alpha} ( p + k_1 )_{\beta}}{p^4 ( p + k_1 )^2} \left[ ( p + k )_{\gamma} - \frac{2 ( k p ) p_{\gamma}}{p^2} \right] + \ldots \; .
\end{equation}
We now use the mixed parameterization $( u, \beta )$, shift the loop momentum $p$ and perform the symmetric integration in the shifted momentum. In the limit $\LL t \to 0$ one can simplify the resulting exponent to
\begin{equation}
- ( \LL t + \beta ) k_1^2  u ( 1 - u ) \; \label{eqn:simplified_exponent_1}
\end{equation}
with $t \equiv t_1 + t_2 + t_3$.
 
After performing the parameter integrals one can exhibit the logarithmically divergent contribution
\begin{eqnarray} 
{\cal W}^{( 3 )}_{\mu \nu \rho} & = & \frac{\imu}{72 {\pi}^2} \biggl\{ \ln (\LL t k_1^2 ) \left[ {\delta}_{\mu \nu} {\left( k_2 - k_1 \right)}_{\rho} - {\delta}_{\nu \rho} {\left( k_1 +  2 k_2 \right)}_{\mu} + {\delta}_{\mu \rho} {\left( 2 k_1 + k_2 \right)}_{\nu} \right]  \nonumber \\ 
                                &    & \mbox{} + \frac{1}{2} {\epsilon}_{\mu \nu \rho \alpha} \left[ \left( 1 - \frac{3 t_1}{t} \right) k_{1 \alpha} - \left( 1 - \frac{3 t_3}{t} \right) k_{2 \alpha} \right] \biggr\} + {\cal O} ( \LL t ) \; . \label{eqn:w3_integrand}
\end{eqnarray}
In getting \eqref{eqn:w3_integrand} we use \eqref{eqn:theorem_integral_zero} or \eqref{eqn:theorem_integral_nonzero} for each covariant. Due to gauge invariance the logarithmic contributions group together in the manner indicated and the factor in front is computed from \eqref{eqn:theorem_integral_nonzero}. The term in the last line is finite but cannot be absorbed in the logarithm. It has been obtained from \eqref{eqn:theorem_integral_zero}.

By the method of section \ref{sec:ward_identity} one can show that the $\epsilon$-term in \eqref{eqn:w3_integrand} does not contribute when inserted into \eqref{eqn:w3} and one makes the identification \eqref{eqn:identification_gauge_field}. Finite contributions are obtained by expanding the logarithm according to \eqref{eqn:expansion_logarithm}. The divergent contribution of the three-leg diagram is
\begin{eqnarray} 
W^{div}_3 ( A ) & = & - \frac{1}{24 {\pi}^2} \int d{\tau}_1 d{\tau}_2 d{\tau}_3  \; \ln ( \LL t {\mu}^2 ) \\
                &   & \mbox{} \times \int d_4 x \tr \left\{ \left[ {\partial}_{\mu} A_{\nu} ( x, {\tau}_1 ) -  {\partial}_{\nu} A_{\mu} ( x, {\tau}_1 ) \right] A_{\mu} ( x, {\tau}_2 )  A_{\nu} ( x, {\tau}_3 ) \right\} \; . \nonumber 
\end{eqnarray}
Proceeding as above one can restrict the $\tau$ integration to the interior of a small cube centered at the origin:
\begin{equation}
| {\tau}_3 \pm {\tau}_1 |, \; | {\tau}_2 \pm {\tau}_1 |, \; | {\tau}_3 \pm {\tau}_2 | \; \leq \frac{T}{3} \; .
 \label{eqn:tau_integration_domain_3}
\end{equation}
Because the $\tau$ integral is invariant under a scaling we can always adjust the linear dimension of the cube to coincide with that of the square, i.~e. the quantity $T$ in \eqref{eqn:tau_integration_domain_2} and \eqref{eqn:tau_integration_domain_3} has the same value. The result is:
\begin{equation} 
W^{div}_3 ( a ) = - \frac{\ln ( \LL T {\mu}^2 )}{24 {\pi}^2} \int d_4 x \tr \left\{ {\partial}_{\mu} a_{\nu} ( x ) \left[ a_{\mu} ( x ) , a_{\nu} ( x ) \right] \right\} \; . \label{eqn:w3_result} 
\end{equation}

The four-leg diagram is given by
\begin{eqnarray} 
W_4 ( A ) & = & \int d{\tau}_1 \ldots d{\tau}_4 \int \frac{d_4 k_1 d_4 k_2 d_4 k_3}{( 2 \pi )^{12}} \; {\cal W}^{( 4 )}_{\mu \nu \rho \sigma} \left( {\tau}_1, \ldots , {\tau}_4 ; k_1, k_2, k_3 \right) \nonumber \\
          &   & \mbox{} \times \tr \left[ A_{\mu} ( - k_1, {\tau}_1 ) A_{\nu} ( - k_2, {\tau}_2 ) A_{\rho} ( - k_3, {\tau}_3 ) A_{\sigma} (  k, {\tau}_4 ) \right] \; , \label{eqn:w4}
\end{eqnarray} 
where 
\begin{eqnarray} 
{\cal W}^{( 4 )}_{\mu \nu \rho \sigma} & = &  - \frac{1}{4} \tr \left( {\gamma}_{\alpha} {\gamma}_{\mu}{\gamma}_{\beta} {\gamma}_{\nu} {\gamma}_{\gamma} {\gamma}_{\rho} {\gamma}_{\delta} {\gamma}_{\sigma} P_L \right) \nonumber \\
                                       &   & \mbox{} \times \int \frac{d_4 p}{( 2 \pi )^4} \frac{p_{\alpha} ( p + k_1 )_{\beta} ( p+ k_1 + k_2 )_{\gamma} ( p + k )_{\delta}}{p^2 ( p + k_1)^2  ( p + k_1 + k_2 )^2 ( p + k )^2} \label{eqn:w4_integrand} \\ 
                                       &   & \mbox{} \times \exp \left\{ - \LL \left[ p^2 t_1 + ( p + k_1 )^2 t_2 +  ( p + k_1 + k_2 )^2 t_3 + ( p + k )^2 t_4  \right] \right\} \; . \nonumber  
\end{eqnarray}
In the expansion of the integrand of \eqref{eqn:w4_integrand} we could keep only the first term. Still it is useful to have some $k$ dependence in the denominator. We propose the following expansion:
\begin{equation}
\frac{p_{\alpha} ( p + k_1 )_{\beta} ( p+ k_1 + k_2 )_{\gamma} ( p + k )_{\delta}}{p^2 ( p + k_1 )^2 ( p + k_1 + k_2 )^2 ( p + k )^2} = \frac{p_{\alpha} p_{\beta} p_{\gamma} p_{\delta}}{p^6 ( p + k_1 )^2} + \ldots \; .  
\end{equation}
The next steps are the mixed parameterization and the symmetric integration over the shifted momentum. As before the exponent can be simplified to \eqref{eqn:simplified_exponent_1} in the limit $\lambda t \to 0$ where now $t \equiv t_1 + \ldots + t_4$. After performing the parameter integrals and expanding the logarithm one obtains the contribution relevant for our purpose  
\begin{equation} 
{\cal W}^{( 4 )}_{\mu \nu \rho \sigma} = \frac{\ln ( \LL t {\mu}^2 )}{96 {\pi}^2} \left( {\delta}_{\mu \nu} {\delta}_{\rho \sigma} + {\delta}_{\mu \sigma}{\delta}_{\nu \rho} - 2 {\delta}_{\mu \rho} {\delta}_{\nu \sigma} -  {\epsilon}_{\mu \nu \rho \sigma} \right) \; . \label{eqn:w4_integrand_evaluated}
\end{equation}
By inserting \eqref{eqn:w4_integrand_evaluated} into \eqref{eqn:w4}, restricting the $\tau$ integration to the interior of a hypercube around the origin:
\begin{equation}
| {\tau}_4 \pm {\tau}_1 |, \; | {\tau}_2 \pm {\tau}_1 |, \; | {\tau}_3 \pm {\tau}_2 |, \; | {\tau}_4 \pm {\tau}_3 | 
\; \leq \frac{T}{4} \; ,
\end{equation}
applying the mean value theorem and making the identification \eqref{eqn:identification_gauge_field} we get the divergent contribution of the four-leg diagram:
\begin{equation} 
W^{div}_4 ( a ) = \frac{1}{48 {\pi}^2} \ln ( \LL T {\mu}^2 ) \int d_4 x \tr \left\{ a_{\mu} ( x ) \left[ a_{\mu} ( x ), a_{\nu} ( x ) \right] a_{\nu} ( x ) \right\} \; . \label{eqn:w4_result} 
\end{equation}

By adding \eqref{eqn:w2_result}, \eqref{eqn:w3_result} and \eqref{eqn:w4_result} one finds a local, gauge invariant expression for the divergent part of the effective action of the conventional theory 
\begin{equation} 
W^{div}( a ) = - \frac{1}{96 {\pi}^2} \ln ( \LL T {\mu}^2 ) \int d_4 x \tr {\left[ f_{\mu \nu} ( x ) \right]}^2 \; , \label{eqn:effective_action_divergent_part} 
\end{equation}
where we used the covariant field strength in Lie algebra valued form
\begin{equation}
f_{\mu \nu} ( x ) \equiv {\partial}_{\mu} a_{\nu} ( x ) - {\partial}_{\nu} a_{\mu} ( x ) + \left[ a_{\mu} ( x ), a_{\nu} ( x ) \right] \; . 
\end{equation} 

In order to have a finite effective action it is sufficient to subtract \eqref{eqn:effective_action_divergent_part} from \eqref{eqn:conventional_effective_action}. Since \eqref{eqn:effective_action_divergent_part} is gauge invariant we have
\begin{equation}
\sigma W ( a ) = \sigma \left[ W ( a ) - W^{div} ( a ) \right] \; \label{eqn:brs_of_effective_action}
\end{equation}
where $\sigma$ is the conventional BRS operator defined by
\begin{equation}
\sigma a_{\mu} ( x ) = {\partial}_{\mu} \omega ( x ) + \left[ a_{\mu} ( x ) , \omega ( x ) \right] \; \qquad \; \sigma \omega ( x ) = - {\omega}^2 ( x ) \; . \label{eqn:conventional_brs_transformation}
\end{equation}
Eq.\ \eqref{eqn:brs_of_effective_action} plays a crucial role in understanding the anomaly. We mention here only two 
consequences:
\begin{enumerate}
\item[(i)] Since $W ( a ) - W^{div} ( a )$ is finite, the right hand side of the Ward identity (the anomaly) is always finite.
\item[(ii)] Because the classical action is gauge invariant the only possible violation of the Ward identity could come from the path integral spinorial measure. This fact has been exploited in \cite{fuj}.
\end{enumerate}


\section{The Anomalous Ward Identity} \label{sec:ward_identity}

\vspace{0.5cm}

We extend the BRS operation on the classical source $A_{\mu} ( x, \tau )$ by introducing a ghost function $\OO ( x, \tau )$ with the properties:
\begin{eqnarray}
\SS A_{\mu} ( x, \tau ) & = & {\partial}_{\mu} \OO ( x, \tau ) + \int d{\tau}' \left[ A_{\mu} ( x, {\tau} - {\tau}' ), \OO ( x, {\tau}' ) \right] \; ; \label{eqn:stochastic_brs_gauge} \\
\SS \OO ( x, \tau )     & = & - \int d{\tau}' \OO ( x, {\tau} - {\tau}' ) \OO ( x, {\tau}' ) \; . \label{eqn:stochastic_brs_ghost}
\end{eqnarray}
One can check the nilpotency ${\SS}^2 = 0$. For consistency we have to assume that $\OO ( x, \tau )$ is homogeneous of degree $- 1$ with respect to $\tau$
\begin{equation}
\OO ( x, \rho \tau ) = {\rho}^{- 1} \OO ( x, \tau ) \; \label{eqn:homogeneity_ghost_field}
\end{equation}   
and to take it of mass dimension $2$. In particular $\SS$ describes the conventional BRS $\sigma$ transformation (cf. 
\eqref{eqn:conventional_brs_transformation}) in the limit
\begin{equation}
A_{\mu} = a_{\mu} ( x ) \delta ( \tau ) \; ; \qquad \; \OO ( x, \tau ) = \oo ( x ) \delta ( \tau ) \; . \label{eqn:identification_ward}
\end{equation}
It is instructive to write \eqref{eqn:stochastic_brs_gauge} and \eqref{eqn:stochastic_brs_ghost} in $k$-space
\begin{eqnarray}
\SS A_{\mu} ( k, \tau ) & = & \imu k_{\mu} \OO ( k, \tau ) + \int d{\tau}' d k' \left[ A_{\mu} ( - k', {\tau} - {\tau}' ), \OO ( k + k', {\tau}' ) \right] \label{eqn:stochastic_brs_gauge_momentum} \\ 
\SS \OO ( k, \tau )     & = & - \int d{\tau}' d k' \OO ( - k', {\tau} - {\tau}')  \OO ( k + k', {\tau}' ) \; .
\end{eqnarray}

The action of $\SS$ on \eqref{eqn:stochastic_effective_action} is written in the form
\begin{equation}
\SS W ( A ) = \sum_{n = 1}^{\infty} U_n ( A ) + \sum_{n = 2}^{\infty} V_n ( A ) \; \label{eqn:expansion_terms_brs} 
\end{equation}
where $U_n ( A )$ and $V_n ( A )$ are associated to the derivative and to the commutator part in 
\eqref{eqn:stochastic_brs_gauge_momentum}, respectively. They are given by
\begin{eqnarray}
U_n ( A ) & \equiv & \int \prod_{j=1}^{n + 1} d{\tau}_j \int \prod_{i=1}^n \frac{d_4 k_i}{( 2 \pi )^4} {\cal U}_{{\mu}_1 \ldots {\mu}_n}^{(n)} \left( {\tau}_1, \ldots , {\tau}_{n + 1} ; k_1, \ldots , k_n \right) \label{eqn:un} \\
          &        & \mbox{} \times \tr \left[ A_{{\mu}_1} ( - k_1, {\tau}_1 ) \ldots A_{{\mu}_n} ( - k_n, {\tau}_n ) \OO ( k, {\tau}_{n + 1} ) \right] \nonumber \\ \nonumber \\
V_n ( A ) & \equiv & \int \prod_{j=1}^{n + 1} d{\tau}_j \int \prod_{i=1}^n \frac{d_4 k_i}{( 2 \pi )^4} {\cal V}_{{\mu}_1 \ldots {\mu}_n}^{(n)} \left( {\tau}_1, \ldots , {\tau}_{n + 1} ; k_1, \ldots , k_{n - 1} \right) \; \label{eqn:vn} \\
          &        & \mbox{} \times \tr \left\{ A_{{\mu}_1} ( - k_1, {\tau}_1 ) \ldots A_{{\mu}_{n - 1}} ( - k_{n - 1}, {\tau}_{n - 1} ) \left[ A_{{\mu}_n} ( - k_n, {\tau}_n ), \OO ( k, {\tau}_{n + 1} ) \right] \right\} \; . \nonumber  
\end{eqnarray}
The corresponding loop integrals have the following expressions:
\begin{eqnarray}
{\cal U}_{{\mu}_1 \ldots {\mu}_n}^{(n)} & \equiv & {\cal U}_{{\mu}_1 \ldots {\mu}_n}^{(n)} \left( {\tau}_1, \ldots , {\tau}_{n + 1} ; k_1, \dots , k_n  \right) \label{eqn:un_integrand} \\
                                        &   =    & \mbox{} {\imu}^n \int \frac{d_4 p}{( 2 \pi )^4} \exp \left( - \LL \sum_{j=1}^{n + 1} p_j^2 t_j \right) \tr \left[ \left( \frac{\slashed{p}_1}{p_1^2} - \frac{\slashed{p}_{n + 1}}{p_{n + 1}^2} \right) {\gamma}_{{\mu}_1} \frac{\slashed{p}_2}{p_2^2} {\gamma}_{{\mu}_2} \ldots \frac{\slashed{p}_n}{p_n^2} {\gamma}_{{\mu}_n} P_L \right] \; \nonumber  \\ \nonumber \\ 
{\cal V}_{{\mu}_1 \ldots {\mu}_n}^{(n)} & \equiv & {\cal V}_{{\mu}_1 \ldots {\mu}_n}^{(n)} \left( {\tau}_1, \ldots , {\tau}_{n + 1} ; k_1, \ldots , k_{n - 1} \right) \label{eqn:vn_integrand} \\
                                        &   =    & \mbox{} - {\imu}^n \int \frac{d_4 p}{( 2 \pi )^4} \exp \left( - \LL \sum_{j=1}^{n} p_j^2 {\bar t}_j \right) \tr \left( \frac{\slashed{p}_1}{p_1^2}{\gamma}_{{\mu}_1} \ldots \frac{\slashed{p}_n}{p_n^2}{\gamma}_{{\mu}_n} P_L \right) \; . \nonumber  
\end{eqnarray} 
We use the abbreviations \eqref{eqn:abbrev_1} -- \eqref{eqn:abbrev_2} with $n$ replaced by $n + 1$. In addition we introduce
\begin{eqnarray}
& {\bar t}_1 \equiv  | {\tau}_n + {\tau}_{n + 1} - {\tau}_1 | \; ;  \;     & {\bar t}_i \equiv t_i \; \; \mbox{for $i = 2, \ldots , \; n - 1$} \; ; \nonumber \\
& {\bar t}_n \equiv | {\tau}_n + {\tau}_{n + 1} - {\tau}_{n - 1} | \; ; \; & {\bar t} \equiv \sum_{i = 1}^n {\bar t}_i \;  .
\end{eqnarray} 
Notice that $V_1 ( A ) = 0$ since the trace of a commutator vanishes. The integrals \eqref{eqn:un_integrand} and \eqref{eqn:vn_integrand} are convergent for $\LL = 0$ when $n \geq 5$. We can set $\LL = 0$, insert them in \eqref{eqn:un} and \eqref{eqn:vn} respectively and make the identification \eqref{eqn:identification_ward}. One gets
\begin{equation}
U_n ( a ) + V_n ( a ) = 0  \; \; \qquad \; \; \mbox{for $ n \geq 5$} \; . \label{eqn:u1}
\end{equation}

If $1 \leq n \leq 4$ the integrals are divergent in the limit $\LL = 0$. We shall now show however that the sum $U_n ( A ) + V_n ( A )$ remains finite in the limit \eqref{eqn:identification_ward}. For $n = 1$ we have 
\begin{eqnarray} 
U_1 ( A ) & = & \int d{\tau}_1 d{\tau}_2 \int \frac{d_4 k}{( 2 \pi )^4} \; {\cal U}_{\mu}^{( 1 )} \left( {\tau}_1, {\tau}_2 ; k \right) \; \nonumber \\
          &   & \mbox{} \times \tr \left[ A_{\mu} ( - k, {\tau}_1 )  \OO ( k, {\tau}_2 ) \right] \; 
\end{eqnarray}
where
\begin{eqnarray}
{\cal U}_{\mu}^{( 1 )} & = & \imu  \int \frac{d_4 p}{( 2 \pi )^4} \exp \left\{ - \LL \frac{t}{2} \left[ p^2 + ( p + k )^2 \right] \right\} \; \nonumber \\
                       &   & \mbox{} \times \tr \left\{ \left[ \frac{\slashed{p}}{p^2} - \frac{\slashed{p} + \slashed{k}}{( p + k )^2} \right] {\gamma}_{\mu} P_L \right\} \; .    
\end{eqnarray} 
Following the way described in section \ref{sec:effective_action} (eqs.\ \eqref{eqn:w2parallel_integrand}, 
\eqref{eqn:w2parallel_integrand_rho}) we obtain the interesting relation
\begin{equation}
{\cal U}_{\mu}^{(1)} = 2 \imu k_{\mu} {\cal W}^{( 2 \; \parallel )} \; .
\end{equation}
By inserting it into \eqref{eqn:u1} and using the homogeneity properties \eqref{eqn:homogeneity_gauge_field} and \eqref{eqn:homogeneity_ghost_field} of both $A_{\mu} ( x, \tau )$ and $\OO ( x, \tau )$ we find that $U_1 (A )$ vanishes. 

A similar method is used for the scalar part of $V_2 ( A )$. The starting point is
\begin{eqnarray} 
V_2 ( A ) & = & \int d{\tau}_1 d{\tau}_2 d{\tau}_3 \int \frac{d_4 k_1 d_4 k_2}{( 2 \pi )^8} {\cal V}^{( 2 )}_{\mu \nu} \left( {\tau}_1, {\tau}_2, {\tau}_3 ; k_1 \right) \nonumber \\
          &   & \mbox{} \times \tr \left\{ A_{\mu} ( - k_1, {\tau}_1 ) \left[ A_{\nu} ( - k_2, {\tau}_2 ), \OO (  k, {\tau}_3 ) \right] \right\} \; , \label{eqn:v2}
\end{eqnarray}
where
\begin{eqnarray}
{\cal V}_{\mu \nu}^{( 2 )} & = & -  \int \frac{d_4 p}{( 2 \pi )^4} \exp \left\{ - \LL \frac{{\bar t}}{2} \left[ p^2 + ( p + k_1 )^2 \right] \right\} \nonumber \\ 
                           &   & \mbox{} \times \tr \left[ \frac{\slashed{p}}{p^2} {\gamma}_{\mu} \frac{\slashed{p} + \slashed{k}}{( p + k_1)^2} {\gamma}_{\nu} P_L \right] \label{eqn:v2_integrand}
\end{eqnarray} 
and ${\bar t} \equiv 2 | {\tau}_2 + {\tau}_3 - {\tau}_1 |$. One can separate \eqref{eqn:v2_integrand} in a scalar and a transversal part:
\begin{equation}
{\cal V}_{\mu \nu}^{( 2 )} = {\delta}_{\mu \nu} {\cal V}^{( 2 \; \parallel )} +  \left( k_{1 \mu}k_{1 \nu} - {\delta}_{\mu \nu} k_1^2 \right) {\cal V}^{( 2 \; \perp )} \; .
\end{equation}
Exactly as before one can show that the integrated version of the scalar part $V_{2 \; \parallel} ( A )$ vanishes.

The remaining integrals ${\cal U}_{\mu \nu}^{( 2 )}$, ${\cal U}_{\mu \nu \rho}^{( 3 )}$, ${\cal V}^{( 2 \; \perp )}$ and ${\cal V}_{\mu \nu \rho}^{( 3 )}$ are expected to be logarithmically divergent in the limit $\LL = 0$. The way to show this is very similar to the method developed in section \ref{sec:effective_action}. We expand the integrand in power of the external momenta $k_i$, introduce a mixed parameterization, perform the loop integral and discard the terms ${\cal O} ( \LL t )$. As a result we obtain a homogeneous polynomial in $k_i$ whose coefficients are expressions logarithmic in $\LL$ and/or $\LL$ independent. In particular ${\cal U}_{\mu \nu \rho \sigma}^{( 4 )}$ turns out to be convergent and vanishing in the limit $\LL = 0$. We shall show later that $V_4 ( a ) = 0$, despite the fact that the corresponding loop integral ${\cal V}^{( 4 )}_{\mu \nu \rho \sigma}$ is logarithmically divergent in the limit $\LL = 0$.

Due to the scaling properties of $A_{\mu} ( x, \tau )$ and $\OO ( x, \tau )$ with respect to the fictitious time $\tau$ most of the $\LL$ independent expressions can be included in the logarithmic divergence. Those finite terms which cannot be absorbed by the logarithm because they have a different $k$-structure form the anomaly. Since as we shall show the logarithmic divergences exactly cancel between $U_n ( A )$ and $V_n ( A )$ only the anomalous terms survive on the right hand side of the Ward identity. Their sum in the limit $\LL = 0$ makes up the chiral anomaly:
\begin{equation}
\sigma W ( a ) = \left[ U_2 ( a ) + V_2 ( a ) \right] + \left[ U_3 ( a ) + V_3 ( a ) \right] 
\end{equation}

To show explicitly the cancelation it is convenient to replace the loop variable $p$ by $p + k_n$.

Let us compute    
\begin{eqnarray} 
U_2 ( A ) & = & \int d{\tau}_1 d{\tau}_2 d{\tau}_3 \int \frac{d_4 k_1 d_4 k_2}{( 2 \pi )^8} {\cal U}^{( 2 )}_{\mu \nu} \left( {\tau}_1, {\tau}_2, {\tau}_3; k_1, k_2 \right) \nonumber \\
          &   & \mbox{} \times \tr \left[ A_{\mu} ( - k_1, {\tau}_1 ) A_{\nu} ( - k_2, {\tau}_2 ) \OO (  k, {\tau}_3 ) \right] \; , \label{eqn:u2}
\end{eqnarray}
where
\begin{eqnarray}
{\cal U}_{\mu \nu}^{( 2 )} & = & \tr \left( {\gamma}_{\alpha} {\gamma}_{\mu} {\gamma}_{\beta} {\gamma}_{\nu} P_L \right) \int \frac{d_4 p}{( 2 \pi )^4} \left[ \frac{( p + l )_{\alpha}}{( p + l )^2} - \frac{( p + k_2 )_{\alpha}}{( p + k_2 )^2} \right] \frac{( p + k )_{\beta}}{( p + k )^2} \nonumber \\ 
                           &   & \mbox{} \times \exp \left\{ - \LL \left[ ( p + k_2 )^2 t_1 +  ( p + k )^2 t_2 + ( p + l )^2 t_3 \right] \right\} 
\end{eqnarray} 
and $l \equiv k_1 + 2 k_2$. The expression in square brackets is expanded as
\begin{eqnarray}
& & \frac{( p + l )_{\alpha}}{( p + l )^2} - \frac{( p + k_2 )_{\alpha}}{( p + k_2 )^2} = k_{\LL} \left( \frac{{\delta}_{\alpha \LL}}{p^2} - 2 \; \frac{p_{\alpha} p_{\LL}}{p^4} \right) \\
& & \mbox{} + \left( l_{\LL} l_{\TT} - k_{2 \LL} k_{2 \TT} \right) \left( 4 \; \frac{p_{\alpha} p_{\LL} p_{\TT}}{p^6} - \frac{{\delta}_{\alpha \LL} p_{\TT} + {\delta}_{\LL \TT} p_{\alpha} + {\delta}_{\TT \alpha} p_{\LL}}{p^4} \right)  + \ldots \; . \nonumber 
\end{eqnarray}
We introduce now the mixed parameterization $( u, \beta )$, shift the integration variable $p$ and perform the symmetric integration. We further simplify the exponent to (cf. \eqref{eqn:simplified_exponent_1})
\begin{equation}
 - B k^2 u ( 1 - u ) = - ( \LL t + \beta ) k^2  u ( 1 - u ) \; . \label{eqn:simplified_exponent_2}
\end{equation}
Moreover, one can discard all ${\cal O} ( \LL t )$ terms resulting from performing the parameter integrals. The remaining terms contain several covariants whose coefficients are evaluated by \eqref{eqn:theorem_integral_zero} or \eqref{eqn:theorem_integral_nonzero}. The result is
\begin{eqnarray}
{\cal U}_{\mu \nu}^{( 2 )} & = & \frac{1}{24 {\pi}^2} \biggl\{ \left[ k_{2 \mu} k_{2\nu} - k_{1 \mu} k_{1 \nu} - \left( k_2^2 - k_1^2 \right) {\delta}_{\mu \nu} \right] \ln ( \LL t k^2 ) \; \nonumber \\
                           &   & \mbox{} + \frac{t_1 - t_3}{2 t} \left( k_{1 \mu} k_{2 \nu} + k_{1 \nu} k_{2 \mu} + {\delta}_{\mu \nu} k_1 k_2 \right) \label{eqn:u2_integrand} \\
                           &   & \mbox{} - \frac{3 ( t_1 + t_3 )}{2 t} {\epsilon}_{\mu \nu \alpha \beta} k_{1 \alpha} k_{2 \beta} \biggr\} + {\cal O} ( \LL t ) \; . \nonumber
\end{eqnarray}
After inserting it into \eqref{eqn:u2} one can use the symmetry under the simultaneous exchange ${\tau}_1 \leftrightarrow {\tau}_2$, $k_1 \leftrightarrow k_2$ and $\mu \leftrightarrow \nu$. The term in the middle line of \eqref{eqn:u2_integrand} gives a vanishing contribution and one obtains
\begin{eqnarray}
U_2 ( A ) & = & - \frac{1}{24 {\pi}^2} \int d{\tau}_1 d{\tau}_2 d{\tau}_3 \int \frac{d_4 k_1 d_4 k_2}{( 2 \pi )^8} \ln( \LL t k^2 ) \left( k_{1 \mu}k_{1 \nu} - k_1^2 {\delta}_{\mu \nu} \right) \nonumber \\
          &   & \mbox{} \times \tr \left\{ \left[ A_{\mu} ( - k_1, {\tau}_1 ), A_{\nu} ( - k_2, {\tau}_2 ) \right] \OO (  k, {\tau}_3 ) \right\} \; \label{eqn:u2_result} \\
          &   & \mbox{} + \frac{r}{16 {\pi}^2} {\epsilon}_{\mu \nu \alpha \beta} \int d_4 x \tr \left[ {\partial}_{\alpha} a_{\mu} ( x ) {\partial}_{\beta} a_{\nu} ( x ) \omega ( x ) \right] \; , \nonumber   
\end{eqnarray}
where the number $r$ is defined by
\begin{equation}
r \equiv \int d{\tau}_1 d{\tau}_2 d{\tau}_3  \delta ( {\tau}_1 ) \delta ( {\tau}_2 ) \delta ( {\tau}_3 ) \frac{| {\tau}_3 - {\tau}_1 | + | {\tau}_3 - {\tau}_2 |}{| {\tau}_3 - {\tau}_1 | + | {\tau}_2 - {\tau}_1 | + | {\tau}_3 - {\tau}_2 |} \; \label{eqn:r_definition}
\end{equation}
It can be evaluated in several steps
\begin{eqnarray}
r & = & 2 \int d{\tau}_1 d{\tau}_2 d{\tau}_3  \delta ( {\tau}_1 ) \delta ( {\tau}_2 ) \delta ( {\tau}_3 ) \frac{| {\tau}_3 - {\tau}_1 |}{| {\tau}_3 - {\tau}_1 | + | {\tau}_2 - {\tau}_1 | + | {\tau}_3 - {\tau}_2 |} \nonumber \\
  & = & 2 \int d{\tau}_1 d{\tau}_2 d{\tau}_3  \delta ( {\tau}_1 ) \delta ( {\tau}_2 ) \delta ( {\tau}_3 ) \frac{| {\tau}_2 - {\tau}_1 |}{| {\tau}_3 - {\tau}_1 | + | {\tau}_2 - {\tau}_1 | + | {\tau}_3 - {\tau}_2 |} \\
  & = & 1 - \int d{\tau}_1 d{\tau}_2 d{\tau}_3 \delta ( {\tau}_1 ) \delta ( {\tau}_2 ) \delta ( {\tau}_3 ) \frac{| {\tau}_2 - {\tau}_1 |}{| {\tau}_3 - {\tau}_1 | + | {\tau}_2 - {\tau}_1 | + | {\tau}_3 - {\tau}_2 |} \; . \nonumber
\end{eqnarray}
The first line is obtained by the exchange ${\tau}_2 \leftrightarrow {\tau}_1$ in the second term of \eqref{eqn:r_definition}, the second line by the exchange ${\tau}_3 \leftrightarrow {\tau}_2$ and the last line is a consequence of the identity $t_1 + t_3 = t - t_2$. It follows 
\begin{equation}
r = \frac{2}{3} \;.
\end{equation}

For computing the transversal part of $V_2 ( A )$ it is convenient to start with
\begin{eqnarray}
{\cal V}_{\mu \nu}^{( 2 )} & = & \tr \left( {\gamma}_{\alpha} {\gamma}_{\mu} {\gamma}_{\beta} {\gamma}_{\nu} 
P_L \right) \int \frac{d_4 p}{( 2 \pi )^4} \frac{( p + k_2 )_{\alpha} ( p + k )_{\beta}}{( p + k_2 )^2 ( p + k )^2} \nonumber \\    
                           &   & \mbox{}\times \exp \left\{ - \LL \frac{\bar t}{2} \left[ ( p + k_2 )^2  +  ( p + k )^2 \right] \right\} \; . 
\end{eqnarray} 
We expand 
\begin{eqnarray}
\frac{( p + k_2 )_{\alpha}}{( p + k_2 )^2} & = & \frac{( p + k_2 )_{\alpha}}{p^2} - \frac{( 2 pk_2 + k_2^2 ) p_{\alpha} + ( p k_2 ) k_{2 \alpha}}{p^4} \nonumber \\
                                           &   & \mbox{} +  \frac{4 ( p k_2 )^2 p_{\alpha}}{p^6} + \ldots \; ,
\end{eqnarray}    
use the mixed parameterization $( u, \beta )$, shift the variable $p$, perform the symmetric integration and simplify the exponent to \eqref{eqn:simplified_exponent_2}. 

Since one is interested in the transversal part one can ignore all ${\delta}_{\alpha \beta}$ terms in the integrand and keep only
\begin{eqnarray}
k_{1 \alpha} k_{1 \beta} & \displaystyle{\frac{\tr \left( {\gamma}_{\alpha} {\gamma}_{\mu} {\gamma}_{\beta} {\gamma}_{\nu} P_L \right)}{( 4 \pi)^2}} & \int_0^1 d u \int_0^{\infty} d \beta \frac{\beta}{B^2} \exp \left[ - B k^2 u ( 1 - u ) \right] \nonumber \\
                         & 
                                         & \mbox{} \times \left[ {\left( 1 - \frac{\LL {\bar t}}{B} \right)}^2 
{\left( u - \frac{1}{2} \right)}^2 - \frac{1}{4} \right] \; . \label{eqn:finding_transversal_part} 
\end{eqnarray}
The transversal combination $k_{1 \mu}k_{1 \nu} - k_1^2 {\delta}_{\mu \nu}$ can be obtained by replacing the factor $k_{1 \alpha} k_{1 \beta}$ in front of the last integral (eq. \eqref{eqn:finding_transversal_part}) by $k_{1 \alpha}k_{1 \beta} + \frac{1}{2} k_1^2 {\delta}_{\alpha \beta}$. Therefore \eqref{eqn:finding_transversal_part} gives directly ${\cal V}^{( 2 \; \perp )}$. After integrating over $\beta$ and $u$ following the instructions of the appendix one gets 
\begin{equation}
{\cal V}^{( 2 \; \perp )} = \frac{1}{24 {\pi}^2} \ln ( \LL {\bar t} k^2 ) + {\cal O} ( \LL t ) \; . 
\label{eqn:v2transversal}
\end{equation}
By inserting \eqref{eqn:v2transversal} into \eqref{eqn:v2} one can adjust ${\tau}_i$ such that ${\bar t} = t$. Taking into account that the scalar part of $V_2 ( A )$ is vanishing one obtains
\begin{eqnarray}
V_2 ( A ) & = & \frac{1}{24 {\pi}^2} \int d{\tau}_1 d{\tau}_2 d{\tau}_3 \int \frac{d_4 k_1 d_4 k_2}{( 2 \pi )^8} \ln( \LL t k^2 ) \left( k_{1 \mu}k_{1 \nu} - k_1^2 {\delta}_{\mu \nu} \right) \nonumber \\
          &   & \mbox{} \times \tr \left\{ A_{\mu} ( - k_1, {\tau}_1 ) \left[ A_{\nu} ( - k_2, {\tau}_2 ), \OO (  k, {\tau}_3 ) \right] \right\} \; . \label{eqn:v2_result}
\end{eqnarray}
By adding \eqref{eqn:u2_result} to \eqref{eqn:v2_result} the logarithmic terms cancel out and we get
\begin{equation}
U_2 ( a ) + V_2 ( a ) = \frac{1}{24 {\pi}^2} {\epsilon}_{\mu \nu \alpha \beta} \int d_4 x \tr \left[ \omega ( x ) {\partial}_{\alpha} a_{\mu} ( x ) {\partial}_{\beta} a_{\nu} ( x ) \right] \; . \label{eqn:uv2_result}
\end{equation}

Let us now compute
\begin{eqnarray}
U_3 ( A ) & = & \int d{\tau}_1 \dots d{\tau}_4 \int \frac{d_4 k_1 d_4 k_2 d_4 k_3}{( 2 \pi )^{12}} {\cal U}^{( 3 )}_{\mu \nu \rho} \left( {\tau}_1, \ldots , {\tau}_4; k_1, k_2, k_3 \right) \nonumber \\ 
          &   & \mbox{} \times \tr \left[ A_{\mu} ( - k_1, {\tau}_1 ) A_{\nu} ( - k_2, {\tau}_2 ) A_{\rho} (  - k_3, {\tau}_3 ) \OO ( k, {\tau}_4 ) \right] \; . \label{eqn:u3}
\end{eqnarray}
The loop integral is given by
\begin{eqnarray}
{\cal U}_{\mu \nu \rho}^{( 3 )} & = & \tr \left( {\gamma}_{\alpha} {\gamma}_{\mu} {\gamma}_{\beta} {\gamma}_{\nu} {\gamma}_{\gamma} {\gamma}_{\rho} P_L \right) \int \frac{d_4 p}{( 2 \pi )^4} \left[  \frac{( p + l )_{\alpha}}{( p + l )^2} - \frac{( p + k_3 )_{\alpha}}{( p + k_3 )^2} \right] \frac{( p + k_{13} )_{\beta} ( p + k )_{\gamma}}{( p + k_{13} )^2 ( p + k )^2} \nonumber \\ 
                                &   & \mbox{} \times \exp \left\{ - \LL \left[ ( p + k_3 )^2 t_1 +  ( p + k_{13} )^2 t_2 + ( p + k )^2 t_3 + ( p + l )^2 t_4 \right] \right\} \; , \label{eqn:u3_integrand}
\end{eqnarray} 
where $k_{13} \equiv k_1 + k_3$ and $l \equiv k_1 + k_2 + 2 k_3$. It is convenient to use the following expansion:
\begin{eqnarray}
& & \left[  \frac{( p + l )_{\alpha}}{( p + l )^2} - \frac{( p + k_3 )_{\alpha}}{( p + k_3 )^2} \right] \frac{( p + 
k_{13} )_{\beta} ( p + k )_{\gamma}}{( p + k_{13} )^2 ( p + k )^2} \nonumber \\
& & \mbox{} = \left[ k_{\alpha} - \frac{2 ( k p ) p_{\alpha}}{p^2} \right] 
\frac{p_{\beta} p_{\gamma}}{p^4 ( p + k )^2}  + \ldots \; .
\end{eqnarray}
By going through all the steps, which are standard by now, one finds
\begin{eqnarray} 
{\cal U}^{( 3 )}_{\mu \nu \rho} & = & \frac{\imu}{24 {\pi}^2} \biggl[ \ln (\LL t k_1^2 ) \left( {\delta}_{\mu \nu} k_{\rho} + {\delta}_{\nu \rho} k_{\mu} - 2 {\delta}_{\mu \rho} k_{\nu} \right) \nonumber \\
                                &   & \qquad\quad - \frac{1}{2} {\epsilon}_{\mu \nu \rho \alpha} k_{\alpha} \biggr] + {\cal O} ( \LL t ) \; .
\end{eqnarray}
Inserting \eqref{eqn:u3_integrand} into \eqref{eqn:u3} one gets
\begin{eqnarray}
U_3 ( A ) & = & \frac{\imu}{24 {\pi}^2} \int d{\tau}_1 \ldots d{\tau}_4 \int \frac{d_4 k_1 d_4 k_2 d_4 k_3}{( 2 \pi )^{12}} \ln ( \LL t k^2 ) \left( {\delta}_{\mu \nu} k_{\rho} + {\delta}_{\nu \rho} k_{\mu} - 2 {\delta}_{\mu \rho} k_{\nu} \right) \nonumber \\ 
          &   & \mbox{} \times \tr \left[ A_{\mu} ( - k_1, {\tau}_1 ) A_{\nu} ( - k_2, {\tau}_2 ) A_{\rho} ( - k_3, {\tau}_3 ) \OO ( k, {\tau}_4 ) \right] \label{eqn:u3_result} \\
          &   & \mbox{} - \frac{1}{48 {\pi}^2} {\epsilon}_{\mu \nu \rho \alpha} \int d_4 x \tr \left[ a_{\mu} ( x )a_{\nu} ( x ) a_{\rho} ( x ) {\partial}_{\alpha}\omega ( x ) \right] \; . \nonumber
\end{eqnarray}

Let us consider
\begin{eqnarray}
V_3 ( A ) & = & \int d{\tau}_1 \ldots d{\tau}_4 \int \frac{d_4 k_1 d_4 k_2 d_4 k_3}{( 2 \pi )^{12}} {\cal V}^{( 3 )}_{\mu \nu \rho} \left( {\tau}_1, \ldots , {\tau}_4; k_1, k_2 \right) \nonumber \\ 
          &   & \mbox{} \times \tr \left\{ A_{\mu} ( - k_1, {\tau}_1 ) A_{\nu} ( - k_2, {\tau}_2 ) \left[  A_{\rho} (  - k_3, {\tau}_3 ), \OO ( k, {\tau}_4 ) \right] \right\} \; , \label{eqn:v3}
\end{eqnarray}
where
\begin{eqnarray}
{\cal V}_{\mu \nu \rho}^{( 3 )} & = & \tr \left( {\gamma}_{\alpha} {\gamma}_{\mu} {\gamma}_{\beta} {\gamma}_{\nu} {\gamma}_{\gamma} {\gamma}_{\rho} P_L \right) \int \frac{d_4 p}{( 2 \pi )^4} \frac{( p + k_3 )_{\alpha} ( p + k_{13} )_{\beta} ( p + k )_{\gamma}}{( p + k_3 )^2 ( p + k_{13} )^2 ( p + k )^2} \nonumber \\
                                &   & \mbox{} \times \exp \left\{ - \LL \left[ ( p + k_3 )^2 {\bar t}_1 + ( p + k_{13} )^2 {\bar t}_2 +  ( p + k )^2 {\bar t}_3 \right] \right\} \; . 
\end{eqnarray} 
The relevant expansion in the integrand amounts to
\begin{equation}
\frac{( p + k_3 )_{\alpha} ( p+ k_{13} )_{\beta} ( p + k )_{\gamma}}{p^4 ( p + k )^2} - \frac{p_{\alpha} p_{\beta} p_{\gamma} p \left( k_1 + 2 k_3 \right)}{p^6 ( p + k )^2} + \ldots  
\end{equation}
with the result
\begin{eqnarray} 
{\cal V}^{( 3 )}_{\mu \nu \rho} & = & \frac{\imu}{24 {\pi}^2} \biggl\{ \ln (\LL {\bar t} k^2 ) \left[ {\delta}_{\mu \nu} {\left( k_2 - k_1 \right)}_{\rho} + {\delta}_{\nu \rho} {\left( 2 k_1 +  k_2 \right)}_{\mu} - {\delta}_{\mu \rho} {\left( k_1 + 2 k_2 \right)}_{\nu} \right] \nonumber \\ 
                                &   & \mbox{} - \frac{1}{2} {\epsilon}_{\mu \nu \rho \alpha} \left[ \left( 1 - \frac{3 {\bar t}_1}{{\bar t}}  \right) k_{1 \alpha} - \left( 1 - \frac{3 {\bar t}_3}{{\bar t}} \right) k_{2 \alpha} \right]  \biggr\} + {\cal O} ( \LL {\bar t} ) \; . \label{eqn:v3_integrand}
\end{eqnarray}
When \eqref{eqn:v3_integrand} is inserted into \eqref{eqn:v3} the second line does not contribute because of the antisymmetry in the exchange $( k_1, {\tau}_1 ) \leftrightarrow  ( k_2, {\tau}_2 )$. Hence after rescaling ${\tau}_i$ such that ${\bar t} = t$ one gets
\begin{eqnarray}
V_3 ( A ) & = & - \frac{\imu}{24 {\pi}^2} \int d{\tau}_1 \ldots d{\tau}_4 \int \frac{d_4 k_1 d_4 k_2 d_4 k_3}{( 2 \pi )^{12}} \ln ( \LL t k^2 ) \left( {\delta}_{\mu \nu} k_{\rho} + {\delta}_{\nu \rho} k_{\mu} - 2 {\delta}_{\mu \rho} k_{\nu} \right) \nonumber \\ 
          &   & \mbox{} \times \tr \left[ A_{\mu} ( - k_1, {\tau}_1 ) A_{\nu} ( - k_2, {\tau}_2 ) A_{\rho} (  - k_3, {\tau}_3 ) \OO ( k, {\tau}_4 ) \right] \; . \label{eqn:v3_result}
\end{eqnarray}

From \eqref{eqn:u3_result} and \eqref{eqn:v3_result} one finds
\begin{equation}
U_3 ( a ) + V_3 ( a ) = \frac{1}{48 {\pi}^2} {\epsilon}_{\mu \nu \rho \alpha} \int d_4 x \tr \left\{ \omega ( x ) {\partial}_{\alpha} \left[ a_{\mu} ( x )a_{\nu} ( x ) a_{\rho} ( x ) \right] \right\} \; . \label{eqn:uv3_result} 
\end{equation}

Finally let us discuss
\begin{eqnarray} 
V_4 ( A ) & = & \int d{\tau}_1 \ldots d{\tau}_5 \int \frac{d_4 k_1 \ldots d_4 k_4}{( 2 \pi )^{16}} {\cal V}^{( 4 )}_{\mu \nu \rho \sigma} \left( {\tau}_1, \ldots , {\tau}_5; k_1, k_2, k_3 \right) \nonumber \\ 
          &   & \mbox{} \times \tr \left\{ A_{\mu} ( - k_1, {\tau}_1 ) \ldots \left[ A_{\sigma} (  k_4, {\tau}_4 ), \OO ( k, {\tau}_5 ) \right] \right\} \; . \label{eqn:v4} 
\end{eqnarray}
The logarithmic divergence comes from
\begin{eqnarray} 
{\cal V}^{( 4 )}_{\mu \nu \rho \sigma} & = &  - \frac{1}{4} \tr \left( {\gamma}_{\alpha} {\gamma}_{\mu}{\gamma}_{\beta} {\gamma}_{\nu} {\gamma}_{\gamma} {\gamma}_{\rho} {\gamma}_{\delta} {\gamma}_{\sigma} P_L \right) \nonumber \\
                                       &   & \mbox{} \times \int \frac{d_4 p}{( 2 \pi )^4} \frac{( p + k_4 )_{\alpha} ( p + k_{14} )_{\beta} ( p+ k_{124} )_{\gamma} ( p + k )_{\delta}}{( p + k_4 )^2 ( p + k_{14})^2  ( p + k_{124} )^2 ( p + k)^2} \label{eqn:v4_integrand} \\ 
                                       &   & \mbox{} \times \exp \left\{ - \LL \left[ ( p + k_4)^2 {\bar t}_1 + ( p + k_{14} )^2 {\bar t}_2 +  ( p + k_{124} )^2 {\bar t}_3 + ( p + k )^2 {\bar t}_4  \right] \right\} \; , \nonumber   
\end{eqnarray}
where $k_{14} \equiv k_1 + k_4$ and $k_{124} \equiv k_1 + k_2 + k_4$. The relevant factor in the integrand (second line of \eqref{eqn:v4_integrand}) is expanded to
\begin{equation}
\frac{p_{\alpha} p_{\beta} p_{\gamma} p_{\delta}}{p^6 ( p + k_1 )^2} + \ldots \; . 
\end{equation}
The integration can be easily done and one obtains
\begin{equation} 
{\cal V}^{( 4 )}_{\mu \nu \rho \sigma} = \frac{1}{12 {\pi}^2} \ln (\LL {\bar t} k^2 ) \left[ \frac{1}{2} \left( {\delta}_{\mu \nu} {\delta}_{\rho \sigma} + {\delta}_{\nu \rho} {\delta}_{\mu \sigma} \right) -  {\delta}_{\mu \rho} {\delta}_{\nu \sigma} \right] + {\cal O} ( \LL t ) \; .
\end{equation}
After inserting the result into \eqref{eqn:v4} one can use the symmetry in $( {\tau}_i, k_i )$ and the trace properties to show that \eqref{eqn:v4} vanishes.

Putting together \eqref{eqn:uv2_result} and \eqref{eqn:uv3_result} we get for the right hand side of \eqref{eqn:expansion_terms_brs} after the identification \eqref{eqn:identification_ward}
\begin{eqnarray}
\sigma W ( a ) & = & \frac{1}{24 {\pi}^2} \, {\epsilon}_{\mu \nu \rho \sigma} \int d_4 x \tr \biggl\{ \omega ( x ) {\partial}_{\sigma} \biggl[ {\partial}_{\mu} a_{\nu} ( x ) a_{\rho} ( x ) \nonumber \\
               &   & \qquad\qquad\qquad\qquad\;\;\;\; + \frac{1}{2} a_{\mu} ( x ) a_{\nu} ( x ) a_{\rho} ( x ) \biggr] \biggr\} \; . \label{eqn:ward_identity}
\end{eqnarray}
Eq. \eqref{eqn:ward_identity} is the celebrated chiral anomaly.

We conclude with some comments on the derivation. The coefficient in front of \eqref{eqn:ward_identity} arose by evaluating the apparently ambiguous fictitious time integral \eqref{eqn:r_definition}. Moreover, we discarded several finite contributions to $\sigma W ( a )$ on a similar ground. We can avoid the computation of ambiguous integrals by using the consistency conditions \cite{weszum}. We first observe that the BRS variation of the effective action is finite and depends only upon the gauge field and its first derivative. Second order derivatives occur only in the compensating logarithmically divergent contributions \eqref{eqn:u2_result} and \eqref{eqn:v2_result} of $U_2 ( A )$ and $V_2 ( A )$. We are now in the position to look for the most general solution of the Wess-Zumino consistency conditions and obtain the chiral anomaly up to a unknown multiplicative constant. The procedure was described for the first time in \cite{mar}, but can be repeated now in a more elegant way by using the BRS operation. The multiplicative constant in front can be obtained by comparision with \eqref{eqn:u3_result}, which does not involve any ambiguous fictitious time integration.


\section{Conclusions}

In this paper we used the bulk quantization as a regularization method for a conventional theory of massless fermions coupled to an external gauge field. The divergent part turned out to be gauge invariant. Moreover, the finite part did not contain any local contribution as could be seen from the second line of \eqref{eqn:w3_integrand}. As a consequence the gauge anomaly has its origin in the finite but nonlocal part of the effective action. 

To compute the gauge anomaly explicitly we regularized the Ward identity by appropriately extending the BRS operation to fictitious time dependent sources. This extension is local in position and fictitious frequency. For completeness we presented the full evaluation of the Ward identity. Some comment at the end of section \ref{sec:ward_identity} allowed us to shorten this calculation considerably.
 
In performing the full evaluation of the divergent part of the effective action, as well as of the Ward identity we encountered ambiguous fictitious time integrals. The ambiguity could be resolved by assuming certain symmetry properties. The procedure is very similar to that used in conventional field theory when the shift parameter of the loop integral is fixed in terms of the external momenta.

Finally, we would like to remark that one loop computation in the bulk quantization of the gauge anomaly is as simple as the traditional point splitting or dimensional reduction.



\acknowledgments

S.~M. thanks Helmuth H\"{u}ffel for suggesting him to read the papers by Baulieu and Zwanziger on bulk quantization. T.~K. thanks the Studienstiftung des deut\-schen Volkes for providing the funding of his education and is very grateful to Prof.~Dahmen for his guidance and directions throughout his whole study and especially regarding the thesis.


\appendix

\section{Appendix}

We use Hermitean Dirac matrices ${\gamma}_{\mu}$ with $\mu = 1, 2, 3, 4$, ${\gamma}_5 \equiv - {\gamma}_1{\gamma}_2{\gamma}_3{\gamma}_4$ and the left chiral projector $P_L \equiv \frac{1}{2} (1 + {\gamma}_5)$. All the traces over products of Dirac matrices can be resolved with the following relations:
\begin{eqnarray}
&& \left\{ {\gamma}_{\mu}, {\gamma}_{\nu} \right\} = 2 {\delta}_{\mu \nu} \; ; \qquad {\gamma}_{\alpha}{\gamma}_{\mu}{\gamma}_{\alpha} = - 2 {\gamma}_{\mu} \; ; \qquad \tr \left( {\gamma}_{\mu} {\gamma}_{\nu} P_L \right) = 2 {\delta}_{\mu \nu} \; ; \nonumber \\
&& \tr \left( {\gamma}_{\mu}{\gamma}_{\nu}{\gamma}_{\rho}{\gamma}_{\sigma} P_L \right) = 2 \left({\delta}_{\mu \nu} {\delta}_{\rho \sigma} - {\delta}_{\mu \rho} {\delta}_{\nu \sigma} + {\delta}_{\mu \sigma} {\delta}_{\nu \rho} - {\epsilon}_{\mu \nu \rho \sigma} \right) \; .
\end{eqnarray} 

In this work we deal with expressions of the form
\begin{eqnarray}
& & \int \prod_{j=1}^{n + 1} d{\tau}_j \int \prod_{i=1}^n \frac{d_4 k_i}{( 2 \pi )^4} {\cal X}_{{\mu}_1 \ldots {\mu}_n}^{(n)} \left( {\tau}_1, \ldots , {\tau}_{n + 1} ; k_1, \ldots , k_n ; \LL \right) \nonumber \\
& & \mbox{} \times \tr \left[ A_{{\mu}_1} ( - k_1, {\tau}_1 ) \ldots A_{{\mu}_n} ( - k_n, {\tau}_n ) \OO ( k, {\tau}_{n + 1} ) \right] \; , 
\end{eqnarray}
where the one loop integrals ${\cal X}_{{\mu}_1 \ldots {\mu}_n}^{(n)}$ are invariant with respect to a scaling of the fictitious time variables
\begin{eqnarray}
& & {\cal X}_{{\mu}_1 \ldots {\mu}_n}^{(n)} \left( \rho {\tau}_1, \ldots , \rho {\tau}_{n + 1} ; k_1, \ldots , k_n ; \frac{\LL}{\rho} \right) \nonumber \\ 
& & \mbox{} = {\cal X}_{{\mu}_1 \ldots {\mu}_n}^{(n)} \left( {\tau}_1, \ldots , {\tau}_{n + 1} ; k_1, \ldots , k_n ; \LL \right) \; . \label{eqn:scaling_X}
\end{eqnarray}
Our aim is to compute these loop integrals in the limit $\LL \to 0$. As a first step we can expand the integrand in an appropriate way, as thoroughly presented in the text.

The parameterization $( u, \beta )$ is introduced by
\begin{eqnarray}
\frac{1}{p^{2M} ( p + k )^2} & = & \frac{1}{( M - 1 )!} \int_{0}^{1} du ( 1 - u )^{M - 1} \int_{0}^{\infty} d \beta {\beta}^M  \nonumber \\
                             &   & \mbox{} \times \exp \left\{ - \beta \left[ p^2 + u ( k^2 + 2 p k ) \right] \right\} \; . 
\end{eqnarray}
After shifting in the momentum variable one can perform the symmetric integration. By using 
\begin{eqnarray}
& & p_{{\alpha}_1} p_{{\alpha}_2} \cdots p_{{\alpha}_{2 n + 1}} \longrightarrow 0 \; ; \qquad p_{\alpha} p_{\beta} \longrightarrow {\delta}_{\alpha \beta} \frac{p^2}{4} \; ; \nonumber \\ 
& & p_{\alpha} p_{\beta} p_{\gamma} p_{\delta} \longrightarrow \left( {\delta}_{\alpha \beta} {\delta}_{\gamma \delta} + {\delta}_{\alpha \gamma} {\delta}_{\beta \delta} + {\delta}_{\alpha \delta} {\delta}_{\beta \gamma} \right) \frac{p^4}{24}
\end{eqnarray} 
the momentum integration can be done explicitly
\begin{equation}
\int \frac{d_4 p}{( 2 \pi )^4} p^{2 N} \exp ( - B p^2 ) = \frac{( N + 1 )!}{( 4 \pi )^2 B^{N + 2}} \; . 
\end{equation}

We are left with a possibly complicated integral over the two parameters $u$ and $\beta$. Being interested in the limit $\LL = 0$ we can simplify at this stage the integrand according to \eqref{eqn:simplified_exponent_1} or \eqref{eqn:simplified_exponent_2}. The resulting integral takes the form
\begin{equation}
I ( \LL ) \equiv \int_0^1 d u \int_1^{\infty} d \xi \frac{P ( u, {\xi}^{- 1} )}{\xi} \exp [ - \LL \xi u ( 1 - u ) ] \; \label{eqn:integral_I}
\end{equation}
where $P ( u, z )$ is a polynomial in $u$ and $z$. On account of \eqref{eqn:scaling_X} one can replace $I ( \LL )$ by $I' ( \LL )$ according to
\begin{equation}
I ( \LL ) \longrightarrow I' ( \LL ) \equiv I \left( \frac{\LL}{\rho} \right) \; . \label{eqn:scaling_I}
\end{equation} 
We shall now prove the following statement:

Under the above assumptions (especially \eqref{eqn:scaling_I}) the value of the integral \eqref{eqn:integral_I} is
\begin{equation}
I' ( \LL ) = \int_0^1 d u \int_1^{\infty} d \xi \frac{P ( u, {\xi}^{- 1} ) - P ( u, 0 )}{\xi} + {\cal O} ( \LL ) \; \label{eqn:theorem_integral_zero} 
\end{equation}
if $\int_0^1 d u P ( u, 0 ) = 0$ and 
\begin{equation}
I' ( \LL ) = - \ln {\LL} \int_0^1 d u P ( u, 0 ) + {\cal O} ( \LL ) \; , \label{eqn:theorem_integral_nonzero}
\end{equation}
otherwise.

For the proof one decomposes \eqref{eqn:integral_I} into two integrals by writing
\begin{equation}
P ( u, {\xi}^{- 1} ) = P ( u, 0 ) + \left[ P ( u, {\xi}^{- 1} ) - P ( u, 0 ) \right] \; .
\end{equation}
In the first term one can perform the $\beta$ integration and show explicitly the logarithmic divergence
\begin{eqnarray} 
\int_1^{\infty} d \xi \exp [ - \LL \xi u ( 1 - u ) ] & = & - \Ei [ - \LL u ( 1 - u )] \nonumber \\ 
                                                     & = & \mbox{} - \gamma - \ln [ \LL u ( 1 - u ) ] + {\cal O} ( \LL ) \; .  
\end{eqnarray}
Here $\Ei ( z )$ denote the integral exponential.

Since the second term contains a polynomial difference it is convergent in the limit $\LL = 0$. The result up to first order in $\LL$ is
\begin{eqnarray}
I ( \LL ) & = & - \int_0^1 d u \left[ \gamma + \ln \LL + \ln u ( 1 - u ) \right] P ( u, 0 ) \nonumber\\
          &   & \mbox{} + \int_0^1 d u \int_1^{\infty} d \xi \frac{P ( u, {\xi}^{- 1} ) - P ( u, 0 )}{\xi} + {\cal O} ( \LL ) \label{eqn:integral_I_result}
\end{eqnarray}
with $\gamma \equiv - {\Gamma}' ( 1 )$ the Euler-Mascheroni constant. Observe that the second integrand behaves as ${\xi}^{- 2}$ at infinity rendering the corresponding integral convergent.

Suppose now the polynomial $P ( u, 0 )$ has the form $\sum_{M = 0}^R c_M u^M$. Then each monomial $c_M u^M$ can be explicitly integrated over $u$. However the whole contribution which is finite for $\LL = 0$ can be discarded by an appropriate rescaling according to \eqref{eqn:scaling_I}. The result is: 
\begin{equation}
\int_0^1 du c_M u^M \int_1^{\infty} d \xi \exp [ - \LL \xi u ( 1 - u ) ] = - c_M \frac{\ln \LL}{M + 1} + {\cal O} ( \LL ) \; . 
\end{equation}

When the total coefficient of $\ln \LL$ is vanishing, i.~e. when we have  
\begin{equation} 
\int_0^1 d u P ( u, 0 ) = - \sum_{M = 0}^R \frac{c_M}{M + 1} = 0 \; \label{eqn:zero}
\end{equation}
the integral \eqref{eqn:integral_I} is convergent at $\LL = 0$ and only the second line of \eqref{eqn:integral_I_result} contributes. Because the logarithmic term disappeared no further rescaling is possible. We obtain \eqref{eqn:theorem_integral_zero}. 

If \eqref{eqn:zero} does not hold, one can perform a new rescaling to get rid of the finite contribution of the second line in \eqref{eqn:integral_I_result}. The result has been announced in \eqref{eqn:theorem_integral_nonzero}. 

In general there will be several polynomials $P ( u, {\xi}^{-1} )$ associated with the various covariants contained in one loop integrals ${\cal X}_{{\mu}_1 \ldots {\mu}_n}^{(n)}$, each of them being treated as above.



\begin{thebibliography}{99}

\bibitem{damhue} P.H. Damgaard and H. H\"{u}ffel: \emph{Stochastic Quantization}, \prep{152}{1987}{227} 

\bibitem{nam} M. Namiki: \emph{Stochastic Quantization}, Springer, Berlin 1992

\bibitem{bauzwa1} L. Baulieu and D. Zwanziger: \emph{QCD(4) from a Five-Dimensional Point of View}, \npb{581}{2000}{604}, {\tt hep-th/9909006}

\bibitem{grabauzwa} L. Baulieu, P. A. Grassi and D. Zwanziger: \emph{Gauge and Topological Symmetries in the Bulk Quantization of Gauge Theories}, \npb{597}{2001}{583}, {\tt hep-th/0006036}  

\bibitem{bauzwa2} L. Baulieu and D. Zwanziger: \emph{From Stochastic Quantization to Bulk Quantization: Schwinger-Dyson Equations and S-Matrix}, \jhep{0108}{2001}{016}, {\tt hep-th/0012103}

\bibitem{sak} B. Sakita: \emph{Stochastic Quantization} in \emph{Proceedings of the Johns Hopkins Workshop on Current Problems in Particle Theory 7}, Bonn 1983, edited by G. Domokos and S. Kovesi-Domokos, World 
Scientific, Singapore 1983

\bibitem{dipl} T. Klose: \emph{Stochastic Quantization of Non-Abelian Gauge Theories and the Computation of the Chiral Anomaly} (Diplomarbeit), Siegen August 2001

\bibitem{oka} K. Okano: \emph{Background Field Method in Stochastic Quantization}, \npb{289}{1987}{109}  

\bibitem{marschokazhe} S. Marculescu, K. Okano, L. Sch\"{u}lke and B. Zheng: \emph{Background Field Method in Stochastic Quantization: Two Loop Renormalization of a Dynamical Yang-Mills Field in Zwanziger Gauge}, \npb{417}{1994}{579}

\bibitem{breit} J. D. Breit, S. Gupta and A. Zaks: \emph{Stochastic Quantization and Regularization}, \npb{233}{1984}{61}

\bibitem{tzani} R. Tzani: \emph{Evaluation of the Chiral Anomaly by the Stochastic Quantization Method}, \prd{33}{1986}{1146}

\bibitem{rogers} A. Rogers: \emph{Supersymmetry and Brownian Motion on Supermanifolds}, {\tt quant-ph/0201006}

\bibitem{wei} S. Weinberg: \emph{The Quantum Theory of Fields}, Vol II, Sect. 22.3, Cambridge University Press 1996

\bibitem{itzzub} C. Itzykson and J.-B. Zuber: \emph{Quantum Field Theory}, Sect. 7.1.1., Clarendon Press, Oxford 1980

\bibitem{fuj} K. Fujikawa: \emph{Path Integral Measure for Gauge Invariant Fermion Theories}, \prl{42}{1979}{1195}

\bibitem{weszum} J. Wess and B. Zumino: \emph{Consequences of Anomalous Ward Identities}, \plb{37}{1971}{95} 

\bibitem{mar} S. Marculescu: \emph{Anomalies for Any Lie Group}, \nc{16A}{1973}{69}

\end{thebibliography}
\end{document}